\documentclass[12pt]{article}
\newcommand{\ket}[1]{\vert#1\rangle}

\newcommand{\ketbra}[2]{\vert#1\rangle\langle#2\vert}

\newcommand{\bd}{\begin{displaymath}}
\newcommand{\ed}{\end{displaymath}}
\newcommand{\be}{\begin{equation}}
\newcommand{\ee}{\end{equation}}
\newcommand{\bq}{\begin{quote}}
\newcommand{\eq}{\end{quote}}
\newcommand{\ben}{\begin{enumerate}}
\newcommand{\een}{\end{enumerate}}
\newcommand{\bi}{\begin{itemize}}
\newcommand{\ei}{\end{itemize}}
\newcommand{\bdes}{\begin{description}}
\newcommand{\edes}{\end{description}}

\begin{document}
\setlength{\baselineskip}{16pt}
\title{AGAINST ``KNOWLEDGE''}
\author{Ulrich Mohrhoff\cite{ujm}\\
Sri Aurobindo International Centre of Education\\
Pondicherry-605002 India}
\date{}
\maketitle 
\begin{abstract}
\normalsize\noindent
Quantum ``states'' are neither states of Nature nor states of knowledge; they are objective 
probability measures. An objective probability measure is the formal expression of an 
objective indefiniteness, such as the positional indefiniteness that contributes to 
``fluff out'' matter. An objective indefiniteness entails that the values of certain 
observables are extrinsic (possessed only because they are indicated) rather than 
intrinsic (indicated because they are possessed). This dependence on value-indicating 
facts is not a dependence on anything external to the quantum world. The latter 
constitutes a free-standing reality that owes nothing to observers, information, 
or our interventions into the course of Nature. The core interpretational issue 
does not concern the kind of reality that quantum states possess---ontological 
or epistemic---but the kind of reality that the spatial and temporal referents 
of quantum states possess. What stands in the way of even the correct identification 
of the central issue is our habit of projecting into the quantum world the detached, 
intrinsically differentiated spatiotemporal background of classical physics.
\setlength{\baselineskip}{14pt}
\end{abstract}

\section{\large OVERVIEW}

The title of this article echoes that of two well-known articles by John 
Bell~\cite{BellAM}. Bell objected to the special role played by measurements in standard 
formulations of quantum mechanics (QM):
\bq
Why this aversion to ``being'' and insistence on ``finding''?

To restrict quantum mechanics to be exclusively about piddling laboratory operations is 
to betray the great enterprise. A serious formulation will not exclude the big world 
outside the laboratory.~\cite{BellAM}
\eq
David Mermin gave vent to the same feeling:
\bq
Why should the scope of physics be restricted to the artificial contrivances we are forced 
to resort to in our efforts to probe the world? Why should a fundamental theory have to 
take its meaning from a notion of ``measurement'' external to the theory itself? Should 
not the meaning of ``measurement'' emerge from the theory, rather than the other way 
around? Should not physics be able to make statements about the unmeasured, 
unprepared world?~\cite{Mermin98}
\eq
In a response~\cite{Mohrhoff00} to Mermin I explained QM's inevitable 
reference to measurements in a way that does not imply a restriction 
of the scope of physics to ``piddling laboratory operations.'' So understood, 
measurements are {\it not\/} external to the theory, the special status of measurements 
{\it does\/} emerge from the theory itself, physics {\it is\/} able to make statements about 
an unmeasured world, even though {\it there is no unmeasured world\/}. Obviously 
two distinct notions of ``measurement'' are involved, and our 
failure to pry them apart is what is largely responsible for our failure to make sense of 
QM.

Bell was right; ``measurement'' is a bad word. It is a bad way of expressing the extrinsic 
nature of such properties as positions or orientations. One of the reasons why Bell 
thought that ``measurement'' was a bad word is that it makes us ``think of the result as 
referring to some {\it preexisting property\/} of the object in question''~\cite[original 
emphasis]{BellAM}. In other words, it makes us think of properties as {\it intrinsic\/} 
(indicated because they are possessed) rather than as {\it extrinsic\/} (possessed because 
they are indicated). The value of a quantum-mechanical observable such as position or a 
spin component is possessed only if, and only to the extent that, it is indicated by, or 
inferable from, a fact---an actual event or state of affairs. {\it No property is a possessed 
property unless it is an indicated property.\/} The relation between possessed positions 
and position-indicating events is obviously best studied in a suitably equipped 
laboratory. But it is just as obvious that this does not imply that position-indicating 
events occur only inside laboratories.

In Sec.~2 I will argue that QM, as a theory of the big world outside the 
laboratory, entails a radically new way of thinking about the world's spatial aspect, in 
which the extrinsic nature of positions plays a crucial role. The special status of 
position-indicating facts emerges from the theory itself, provided that this is taken 
seriously as an {\it ontological\/} theory.

It is much easier, however, to construe QM as an epistemic theory, inasmuch as this 
saves us from the necessity of learning a new way of thinking about space. Conversely, if 
we persist in viewing the quantum world against the self-existent and intrinsically 
differentiated spatial background of classical physics, it seems impossible to avoid the 
conclusion that
\bq
[t]here is no quantum world. There is only an abstract quantum physical description. It 
is wrong to think that the task of physics is to find out how nature is. Physics concerns 
what we can say about nature.~\cite{Petersen}
\eq
In his response to Bell, Rudolf Peierls likewise defended the view that
\bq
the most fundamental statement of quantum mechanics is that the wave function or, 
more generally the density matrix, represents our {\it knowledge\/} of the system we are 
trying to describe.~\cite[original emphasis]{Peierls}
\eq
Since the beginning of time (in about 1926) it has been argued that quantum theory is 
about knowledge or, as they now say, about 
information~\cite{LonBau,vN,Wigner,Heisbg}. There are various reasons for this 
attitude. For one thing, it is safe. I can't prove it wrong, as little as I can disprove the 
claim that the world was created yesterday with all our memories in place. (In this case 
we would have knowledge of a nonexistent past, rather than knowledge of a nonexistent 
world.) Inasmuch as it is irrefutable, this attitude is also 
unscientific, at least according to Karl Popper's definition of ``science''~\cite{Popper}. 
For another thing, it seems obvious. If the properties of the quantum world are extrinsic 
(that is, if they ``dangle'' from, or supervene on, something), and if the quantum world 
is coextensive with the physical world, then from what ``hook'' can they ``dangle''? The 
obvious answer is, from us, from what we perceive, or from what we know.

But is it true that the quantum world is coextensive with the physical world? A.~Peres 
and W.H. Zurek have forcefully argued against it: ``[A]lthough it can describe {\it 
anything\/}, a quantum description cannot include {\it everything\/}''~\cite[original 
emphasis]{PerZur}. Yet this by itself does not save us from the conclusion that the wave 
function represents knowledge; it rather appears to confirm it. Bell demanded that ``the 
theory should be fully formulated in mathematical terms, with nothing left to the 
discretion of the theoretical physicist''~\cite{BellAM}. What appears to be the case, 
instead, is that the division of the world into ``system'' and ``apparatus'' is both 
arbitrary and inevitable. This suggests that the ``full elision of the 
subject''~\cite{Bitbol} cannot be achieved: We, the creators of the theory, are unable to 
entirely withdraw from what the theory describes~\cite{dESEP}. Because the description 
carries our signature---the ``shifty split'' between system and apparatus---it must not be 
mistaken for the thing described.
\bq
There can be no question\dots without changing the axioms\dots of getting rid of the 
shifty split. Sometimes some authors of ``quantum measurement'' theories seem to be 
trying to do just that. It is like a snake trying to swallow itself by the tail. It can be 
done\dots up to a point. But it becomes uncomfortable for the spectators even before it 
becomes painful for the snake.~\cite[original ellipses]{BellAM}
\eq
Bell's simile of a snake swallowing its own tail is appropriate. Position-indicating 
positions, such as the position of a detector or a pointer needle, are extrinsic, too; they 
too need to be indicated in order to be possessed. This seems to send us chasing the 
ultimate property-defining facts in never-ending circles. But only {\it seems\/}. In truth 
there is a perfectly natural and perfectly objective way of dividing the physical world 
into a macroscopic and a quantum domain. The divide is {\it not\/} left 
to the discretion of the physicist. It {\it emerges\/} from the theory. The ``full elision of 
the subject'' can be achieved. {\it But not without a radical change in our 
conception of physical space. Not if we insist on projecting the self-existent and 
intrinsically differentiated spatial background of classical physics into the physical 
world.\/} What makes that exercise painful to watch is the use of an inappropriate 
concept of space. This is the subject of Sec.~3.

In Bell's view~\cite{BellAM}, QM, as taught in most textbooks, is only {\it for all 
practical purposes\/} (FAPP) consistent with the ideal of an objective description in 
which nothing is left to the discretion of theoretical physicists. A classic case in point, 
quoted by Bell, is Kurt Gottfried's justification of the substitution of the mixture
\bd
\hat{\rho}=\sum_n|c_n|^2\ketbra{\psi_n}{\psi_n}
\ed
for the pure state
\bd
\rho=\sum_m\sum_nc_m^{*}c_n\ketbra{\psi_m}{\psi_n}:
\ed
``[W]e are free to replace $\rho$ by $\hat{\rho}$ after the measurement, safe in the 
knowledge that the error will never be found''~\cite{Gottfried}. In the same way Asher 
Peres disposes of the difference between $\rho$ and $\hat{\rho}$ by arguing that if it 
were ever observed, it would be considered as a statistical quirk or an 
experimental error~\cite{Peres00}. If safety from the 
consequences of errors in a description of the world is what justifies the 
description then obviously the description is valid only FAPP. Bernard 
d'Espagnat~\cite{dESEP,dERP,dEVR} concludes from this that standard 
QM, although consistent with a weak notion of objectivity based on intersubjective 
agreement, is inconsistent with a strongly objective description free from traces left 
by its creators. Recently Christopher Fuchs and Asher Peres~\cite{FuPer} have argued 
along similar lines that there exists no free-standing reality independent of our 
interventions 
into the course of Nature. There only exists an ``effective reality'' (their shudder quotes) 
that QM ``produces in some regimes of our experience.'' This ``forms the ground for all 
our other quantum predictions simply because it is the part of nature that is effectively 
detached from the effect of our experimental interventions''~\cite{FuPer2}.

In point of fact, the ``\,`effective reality'\,'' of Fuchs and Peres (my shudder quotes 
around theirs) is a perfectly free-standing reality, in the sense that the totality of 
macroscopic positions and the properties these indicate in no wise depend on observers, 
their information, their arbitrary decisions, or their interventions into the course of 
Nature. What this reality is effectively (rather than absolutely) detached from is 
property-indicating facts. 
Positions are extrinsic. While macroscopic objects are no exception, their indicated 
positions are so correlated that they are consistent with definite trajectories. For this 
reason the positions of macroscopic objects are effectively detached from the facts by 
which they are indicated. They can therefore consistently be regarded as intrinsic, or 
self-indicating, or factual {\it per se\/}, and hence as capable of indicating properties, 
including each other.

This conclusion is valid not just FAPP but strictly. It is true that 
arguments based on decoherence~\cite{JoosZeh} and environment-induced 
superselection~\cite{Zurek82} do not lead to strictly vanishing off-diagonal terms, and 
therefore cannot justify the substitution of $\hat{\rho}$ for $\rho$---except FAPP. But it 
is irrelevant. 
What matters is not that the probability of a departure from a classical trajectory is so 
low that it is unlikely that one will ever be observed. What matters is that {\it none is 
ever indicated\/}. This forms part of the definition of ``macroscopic object'' given in 
Sec.~3, where it is also shown that macroscopic objects exist.

To settle the question of what the laws of QM are {\it about\/}, we must be clear what 
the laws of QM {\it are\/}. As I explained in a previous article~\cite{Mohrhoff00}, they 
are concise encapsulations of statistical correlations---diachronic correlations between 
facts concerning the properties of, or the values of observables on, the same system
at different times, and synchronic correlations between facts concerning the properties 
of different systems, or the values of (locally measurable) observables, in spacelike 
separation. Quantum ``states,'' accordingly, are probability measures. They are 
uniquely determined ways of assigning probabilities in the presence of an objective 
indefiniteness, such as the objective positional indefiniteness that is essential for the 
stability of extended material objects. The proper formal expression of an objective 
indefiniteness is to make counterfactual probability assignments, and the search for a 
suitable probability algorithm leads straight to a unique density operator and the 
well-known trace rule. If you were looking for a duck and you found something that 
walks like a duck and that quacks like a duck, you would think that you have found a 
duck. Yet when it comes to quantum ``states,'' many physicists are willing to suspend 
this sensible maxim and attempt to construe an obvious probability algorithm as if it 
represented an actual state of affairs. I am in complete agreement with Fuchs and Peres 
when they criticize such attempts~\cite{FuPer}:
\bq
Attributing reality to quantum states leads to a host of `quantum paradoxes.' These are 
due solely to an incorrect interpretation of quantum theory\dots. The time dependence 
of the wave function does not represent the evolution of a physical system.
\eq
Fuchs and Peres overshoot the mark when they claim that quantum {\it theory\/} (as 
against the quantum {\it state\/} of a system) does not describe physical reality, that the time 
dependence of the wave function gives the evolution of {\it probabilities\/}, and that 
quantum states are states of {\it knowledge\/}. These claims are rooted in the same 
misconception that 
makes other physicists look for ways of construing the density operator as describing 
an actual state of affairs. They are different symptoms of the same illness, diagnosed in 
Sec.~5 as the endemic misconception that quantum states evolve in a detached, 
intrinsically differentiated temporal background. If I were to vote on the worst words in 
the vocabulary of theoretical physics, I would vote for the mutual implicates ``state'' 
and ``evolution.'' Eliminating these words (along with the misconceptions they connote) 
rids us not only of the spurious ``measurement problem''---Why are there two modes of 
evolution?---but also of the false disjunction of possible interpretations of quantum 
states into ontological and epistemic ones. Quantum states are neither states of Nature 
nor states of knowledge. They are generic probability measures.

But does not the presence of nontrivial probabilities in a fundamental theory by itself 
imply that QM is about knowledge? Not any more than the presence of mathematical 
laws in any physical theory does. If QM is about knowledge then so is classical physics. 
Contrariwise, if classical physics is a theory from which a 
free-standing reality can be distilled, as Fuchs and Peres affirm, then QM too is such a 
theory. The statistical laws of QM are as objective as the deterministic laws of classical 
physics. This, along with the question of the completeness of QM, is discussed in Sec.~6.

The concluding section is cast in the form of a comment on a recent review article by 
F. Lalo\"e~\cite{Laloe}.

\section{\large SPACE AND THE QUANTUM WORLD}
\label{SAQW}
If it is to make sense as a theory of the big world outside the laboratory, QM needs a 
radically new way of thinking about the world's spatial 
aspect~\cite{Mohrhoff00,WATQM,SQW}. Space cannot be something that exists by 
itself and that is intrinsically differentiated or divided. Here is why: On the one hand, 
QM tells us that an electron can go through two slits without going through either slit in 
particular and without being divided by the slits~\cite{bucky}. (The present article is 
concerned with the interpretation of standard QM unadulterated with, e.g., Bohmian 
trajectories~\cite{Bohm} or nonlinear modifications of the dynamics~\cite{GRW}.) On 
the other hand, for reasons that are neurophysiological rather than 
physical~\cite{SQW,BCCP,QMCCP}, we tend to think of space as something that exists 
by itself, independently of its material ``content,'' and that is intrinsically divided into 
distinct, separate regions. Yet if this were the case, it would affect every material object. 
No particle could go through $L\&R$---the regions $L$ and $R$ defined by the slits 
considered as one region---without going through a particular slit {\it and\/} without 
being divided into two distinct parts. Therefore $L$ and $R$ cannot be self-existent and 
intrinsically distinct ``parts of space.'' {\it A fortiori\/}, what we conceive of as the 
``parts of space'' cannot be real and distinct {\it per se\/}. A ``part of space'' has a {\it 
contingent\/} reality, in the sense that it may exist for one material object at one time 
and not exist for another object at the same time or for the same object at another time.

This calls for a criterion, and the criterion that suggests itself to me is this: A region of 
space~$V$ is real for an object~$O$ at a time~$T$ if---and only if---two conditions are 
satisfied: (i)~$V$ must exist as an intrinsic property of a macroscopic object~$M$ (to be 
rigorously defined in the next section), and (ii)~the proposition ``$O$~is in~$V$ at the 
time~$T$''---symbolically, $O{\rightarrow}V(T)$---must have a truth value. [In this 
symbolic expression ``$O$'' stands for the {\it position\/} of~$O$, and ``$V$'' stands for 
a region that is realized (made real) by being an intrinsic property of~$M$. The lack of 
a truth value means that the proposition is neither true nor false but meaningless.] And 
the sufficient and necessary condition for the existence of a truth value is that one is 
indicated.

Some comments are in order. Crucial to the proposed interpretational scheme is the 
extrinsic nature of the values of quantum observables. Being essentially a probability 
algorithm, QM presupposes events (specifically, value-indicating events) to which 
probabilities can be assigned. In agreement with Niels 
Bohr I maintain that a value-indicating event or state of affairs is not only sufficient but 
also necessary for the 
existence of a value. But it does not have to be the click of a laboratory counter or the 
deflection of a pointer needle. {\it Any\/} actual event or state of affairs from which the 
possession of a property or a value, by a system or an observable, can in principle be 
inferred, is sufficient for the existence of that property or value~\cite{Mohrhoff00}.

To avoid a potential misunderstanding, let us assume that the presence of $O$ in $V$ is 
indicated. Then the absence of $O$ from the complement $V'$ of $V$ is also indicated. 
If nothing indicates the presence of $O$ in any region $W\subset V$ then condition~(ii) 
is not met for $W$---such a region does not exist for~$O$. The absence of $O$ from any 
region $U\subset V'$, on the other hand, seems to be inferable from the absence of $O$ 
from~$V'$, so that the proposition $O{\rightarrow}U(T)$ seems to have a truth value. 
Suppose it has a truth value but condition~(i) is not satisfied. Then this truth value---the 
truth or falsity of $O{\rightarrow}U(T)$---concerns something that exists in our 
imagination, rather than something that is the case in the real world. If 
$O{\rightarrow}U(T)$ or its negation is to be a meaningful statement about the physical 
world, the region $U$ must exist in the physical world, and for this it must be 
intrinsically possessed (that is, possessed by a macroscopic object, as defined in the 
following section). This is why the region referred to by condition~(ii) is stipulated to be 
an intrinsically possessed region. If $U$ is not realized as an intrinsic property of a 
macroscopic object, the property of being in $U$ is not available for attribution to~$O$. 
It does not form part of physical reality. Hence its possession by $O$ cannot be 
indicated, and it cannot be real for~$O$.

Nobody is likely to assert that red, or a smile, can exist without a red object or a smiling 
face (the smile of the Cheshire cat notwithstanding). Yet, for the neurophysiological 
reasons alluded to, we tend to think that positions (points or regions) can exist 
even when they are not properties of material objects. This way of thinking appears to 
me to be as inconsistent with QM (as a theory about the big world outside the 
laboratory) as absolute simultaneity is with special relativity. In the world 
according to QM~\cite{WATQM}, no position exists unless it is possessed, and no 
position is possessed unless its possession is indicated~\cite{Relationalism}, even though 
no position can be indicated unless it is possessed by a macroscopic object. (The resolution 
of this apparent vicious circle will concern us in the following section.)

When dealing with extrinsic properties a clear distinction has to be made between the 
times of property- or value-indicating events and the times at which the indicated 
properties or values are possessed. This allows for an innocuous kind of advanced 
(backward-in-time) influence, inasmuch as the value-indicating event may happen after 
the time at which the indicated value is possessed~\cite{Mohrhoff00,Mohrhoff99}.

In the context of a two-slit experiment the relevant events are the emission of an 
electron~$e$ in 
front of the slit plate~$S$ and its detection behind~$S$. Together with the geometry of 
the setup they warrant the truth of $e{\rightarrow}L\&R(T)$, while (in the absence of 
any further matter of fact about the electron's whereabouts in the interim) the 
propositions $e{\rightarrow}L(T)$ and $e{\rightarrow}R(T)$ lack truth values. Hence 
whenever it was that the electron went through~$S$, $L\&R$ was real for the 
electron at that time, while neither $L$ nor $R$ were real for it.

But when was that? What is the value of $T$ in the true proposition\break
$e{\rightarrow}L\&R(T)$? 
Just as any material object is spatially localized only to the indicated extent, so an event 
like the electron's passage through~$S$ is temporally localized only to the indicated extent. 
Not only the values of quantum observables are extrinsic but also the times at which they 
are possessed. The extent to which~$T$ is indicated is defined by the respective times of 
emission and detection, $T_e$ and~$T_d$. Just as the electron's position at~$T$ is the 
entire region $L\&R$, so $T$ is the entire interval [$T_e$,$T_d$]. Nothing warrants 
the inference that the electron passed~$S$ at some {\it particular\/} time during this 
interval. All that QM permits us to say in this case is counterfactual and probabilistic: 
Imagine the interval [$T_e$,$T_d$] divided into intervals~$T_i$. If there {\it were\/} 
an event from which the electron's passage through~$S$ during a particular interval~$T_i$ 
could be inferred, it {\it would\/} indicate the interval during which the electron went 
through~$S$ with probability~$p_i$. This set of counterfactuals is the proper formal 
expression of an objective temporal indefiniteness.

\section{\large A FREE-STANDING REALITY}

Even position-indicating positions are extrinsic; they too need to be indicated in order to 
be possessed. To avoid a vicious regress one must show that {\it some\/} positions can 
consistently be thought of as intrinsic, as self-indicating, or as factual {\it per se\/}. Here 
is how this may be done.

There are objects whose indicated positions are so correlated that every one of them is 
consistent with every prediction that is based on (i)~previous indicated positions and 
(ii)~a classical law of motion (except, of course, when their indicated positions 
themselves serve to indicate unpredictable values). If we take this characterization as a 
definition of ``macroscopic object'' then what needs to be shown is that such objects 
exist. Note that this definition does not require that the probability of finding a 
macroscopic object where classically it could not be, is strictly~0. What it requires is that 
there be no position-indicating fact that is inconsistent with predictions based on a 
classical law of motion and earlier position-indicating facts.

The departure of an object $O$ from a classical trajectory can be indicated only if there 
are detectors whose positions are sharper than $O$'s position, or whose position 
probability distributions are narrower than that of~$O$. One of the things QM tells us 
is that the relative position of two objects cannot be exact. Some relative positions are 
sharper than others. Some objects have the sharpest positions in existence. For such 
objects the probability of a position-indicating event that is inconsistent with a classical 
trajectory is necessarily very low. It is therefore certain that {\it among\/} such 
objects there will be macroscopic objects.

Since no object has an exact position, it might be argued that even for a macroscopic 
object~$M$ there always exists a small enough region $V$ such that $M{\rightarrow}V$ 
lacks a truth value. But this is an error. Macroscopic objects have the sharpest positions 
in existence. There isn't any object that has a (significantly) sharper position. {\it A 
fortiori\/}, there isn't any object for which $V$ is real. But a region exists only if it is real 
for at least one material object. It follows that there exists no sufficiently small region 
$V$ such that $M{\rightarrow}V$ lacks a truth value. Such a region may exist in our 
imagination but it does not exist in the physical world.

Now recall why positions are extrinsic: The proposition $O{\rightarrow}V$ may or may 
not have a truth value. One therefore needs a criterion for the existence of a truth value: 
A truth value must be indicated. On the other hand, one doesn't need a criterion for the 
existence of a truth value if {\it for every existing region\/} $V$ the proposition 
$M{\rightarrow}V$ has a truth value. Since this condition is satisfied by macroscopic 
objects, the positions of such objects can consistently be regarded as intrinsic, or factual 
{\it per se\/}.

The laws of classical mechanics define a set of nomologically possible worlds. They do 
not single out the actual world. While exactly one possible world has the mysterious 
quality of being {\it real\/}, classical mechanics cannot tell us which one it is. It has 
no symbol for the property of being real; reality is not a classical observable. QM on the 
other hand seems to possess a criterion of reality: To be is to be indicated. Yet 
being indicated is not a physical observable either. There is no projector or subspace 
that represents the property of being indicated. Instead QM gives us a choice of two 
rules: To calculate the probability of a state of affairs $\cal A$ that can come about in 
several ways, add the probabilities/amplitudes associated with the alternatives if 
something/nothing indicates the alternative taken. (We also 
add probabilities if the various ways in which $\cal A$ can come about 
are so correlated with a set of possible, mutually exclusive events that the occurrence 
of one of these events would indicate the actual way in which $\cal A$ comes about.) QM 
seems to leave it to our discretion whether or not the alternative is indicated.

Let us take a closer look at what seems to be left to our discretion. An electron's 
probability of being detected at a particular location behind the slit plate depends on two 
amplitudes, one for each slit. We add the absolute squares of these amplitudes, rather 
than the amplitudes themselves, if the individual regions $L$ and $R$ are real for the 
electron $e$, or if the propositions $e{\rightarrow}L$ and $e{\rightarrow}R$ have truth 
values, or if their truth values are indicated. The reality, for the electron, of a 
particular region---say, $R$---requires two things: $R$ must exist (that is, it must 
be real for at least one object), and the electron's presence in or absence from $R$ 
must be indicated. A detector---in the broadest sense of the word---serves both purposes: 
It warrants the reality of~$R$, and it indicates the truth value of $e{\rightarrow}R$. 
QM thus presupposes detectors but seems to leave it to our discretion to decide what is, 
and what is not, a detector.

The problem of granting detector status is essentially the problem of identifying 
positions that can be consistently regarded as factual {\it per se\/}. This must not be 
confused with the spurious problem of the ``emergence'' of facts. Factuality 
is not a physical observable; it cannot be measured. The notion that physics ought to 
account for the actuality of the actual world or the factuality of facts is therefore 
misconceived. A theory that describes the actual world without fully determining it 
cannot possibly do this, and even if we had a theory that determined a unique world it 
could not account for the existence of that world, inasmuch as it could not account for its 
own truth. The actual world does not exist by courtesy of laws that describe it. To 
``explain why events occur''~\cite{Pearle79} or ``how it is that probabilities become 
facts''~\cite{Treiman} is as impossible as to explain why there is anything rather than 
nothing.

QM concerns the {\it transmittal\/} of factuality from self-existent, 
property-indica\-ting positions to indicated properties. The philosophically correct term is 
{\it supervenience\/}: Indicated properties supervene 
on indicating properties. Factuality is always presupposed, not only because QM assigns 
probabilities on the basis of facts but also because it assigns them to properties indicated 
by facts. Pointers do deflect, as certainly as the world exists, and just as inexplicably. 
What is inexplicable is not only why the pointer deflects to one side rather than the other 
but also why it deflects at all. QM honors the impossibility of explaining the latter by 
assigning probabilities {\it on 
the condition\/} that the pointer deflects. (The probability that the observable $Q$ has 
the value $q$ is the product of two probabilities: The probability that any one of the 
possible values of $Q$ is indicated, and the probability that the indicated value is~$v$, 
given that a value is indicated. The probabilities QM allows us to calculate are 
exclusively of the latter type.)

For the same reason that classical mechanics does not determine which possible world 
corresponds to the actual one, QM does not determine which possible outcome 
corresponds to the actual one. (``Possible'' here means ``consistent with the laws of 
physics.'') To expect otherwise is to ignore the fundamental indefiniteness that finds 
expression in the statistical laws of QM. A macroscopic pointer, however, is exempt from 
this indefiniteness, inasmuch as its own indicated positions before and after a 
value-indicating deflection are predictably correlated and therefore effectively intrinsic. 
This makes the deflection an unpredictable change from one actual state of affairs to 
another. It is therefore impossible for a macroscopic pointer to ``catch'' the 
indefiniteness of a quantum-mechanical observable. Indefiniteness is not contagious; 
factuality is.

Indefiniteness means that certain propositions lack truth values, and this, we have seen, 
entails the supervenience of indicated properties on indicating properties. There is a dependent 
actuality, and this entails the existence of an independent actuality. What seems to be 
left to our discretion is the attribution of this independent actuality. But only {\it 
seems\/}, for QM itself furnishes the sufficient and necessary criterion of self-existence. 
The positions of macroscopic objects---and only these---can consistently be considered 
factual {\it per se\/}.

The crux of the matter is that space is not something that exists by itself and is 
intrinsically differentiated. Regions of space exist only as properties of matter, to the 
extent that propositions of the form $O{\rightarrow}V$ possess truth values, and a 
region $V$ is real only for those objects $O$ for which $O{\rightarrow}V$ possesses a 
truth value. Since no exact position is ever indicated, infinitesimal regions of space do 
not exist. Nor do sharply bounded regions. The extent to which the world is 
differentiated spacewise is both relative (for some objects greater, for some objects less) 
and finite. For a macroscopic object the world possesses the highest degree of spatial 
differentiation, and the space over which its position is ``smeared out'' is never probed. 
This space is undifferentiated; it contains no smaller regions. We may imagine smaller 
regions, but they don't exist in the physical world.

It is therefore unnecessary to treat the position of a macroscopic object as a probability 
distribution. Probability of what? Nothing in the realm of fact indicates a departure 
from a definite trajectory. The fuzziness of the trajectory only exists in relation to an 
imaginary spatial background that is more differentiated than the actual world. We 
do need to acknowledge that even the positions of macroscopic objects exist only because 
they are indicated---by the positions of macroscopic objects. Each macroscopic position 
presupposes the others; none exists without the others; each has the value that is 
indicated by the others. But the entire system of macroscopic positions is self-contained. 
Together with all the properties that supervene on it, it constitutes a free-standing 
reality---a quantum world that does not depend on observers, their information, 
their arbitrary decisions, or their interventions in ``the course of Nature.'' The 
relevant dependence is the mutual dependence of the positions of macroscopic objects 
and the supervenience, on these, of all possessed properties. This is an {\it internal\/} 
dependence---it does not presuppose anything external to the quantum world.

\section{\large QUANTUM ``STATES'' ARE PROBABILITY\\
MEASURES}

A typical but didactically disastrous approach to QM 
begins with the observation that in classical physics the 
state of a system is represented by a point $\cal P$ in some phase space, and that 
the system's possessed properties are represented by the subsets containing~$\cal P$. 
The question then is, what are the quantum-mechanical counterparts to $\cal P$ and the 
subsets containing~$\cal P$, respectively, {\it qua representations of an actual state of 
affairs and possessed properties\/}? Once we accept this as a valid question, we are on a 
wild-goose chase. It would be better to begin with the observation that in classical 
physics a system is associated with a probability measure, that this is represented by a 
point~$\cal P$ in some phase space, that observable properties are represented by 
subsets, and that the probability of finding a property is~1 if the corresponding set 
contains~$\cal P$; otherwise it is~0. The question then is, what are the 
quantum-mechanical counterparts to $\cal P$ and the subsets containing~$\cal P$, 
respectively, {\it qua representations of a probability measure and observable 
properties\/}?

Classical probability measures assign trivial probabilities: either 0 or~1. This permits us 
to treat $\cal P$ without further ado as representing an actual state of affairs connoting 
a set of possessed properties. Quantal probability measures generally assign 
nontrivial probabilities, and this forbids us to treat them as representing states 
connoting possessed properties. With one exception: For a macroscopic object $M$ 
there exists no region $V$ such that the proposition $M{\rightarrow}V$ lacks a truth 
value. (In this proposition, recall, ``$M$'' stands for the position of $M$.) The position 
probability measure associated with $M$ assigns trivial probabilities to all regions 
of space that can legitimately be said to exist in the physical world. This permits us to 
represent the probability measure associated with $M$ by a point in a phase space, and to 
treat this point as representing a state connoting a set of possessed properties. If this 
point is fuzzy, it is so only relative to a spatial background that is more differentiated 
than the physical world.

Whence the nontrivial probabilities? It is well known that the objective indefiniteness of 
relative positions is crucial for the stability of extended material 
objects~\cite{Lieb}. Together with the exclusion principle it ``fluffs out'' 
matter. The proper way of dealing with objectively indefinite values is to make 
counterfactual probability assignments~\cite{Mohrhoff00,Mohrhoff01}. If a quantity is 
said to have an ``indefinite value,'' what is really intended is that it does not 
have a value (inasmuch as no value is indicated) but that it {\it would\/} have a 
value if one {\it were\/} indicated, and that at least two possible values are associated 
with positive probabilities. Nontrivial probabilities thus are a consequence of the 
objective positional indefiniteness to which extended matter owes its stability.

To find the quantum counterpart to $\cal P$, we have to make room for nontrivial 
probabilities, and the obvious way to do this is to replace the subsets of a phase space by 
the subspaces of a vector space, or by the corresponding projectors. (Why this vector 
space has to be complex will be discussed in a separate article.) An atomic probability 
measure will then be a 1-dimensional subspace~$U$ instead of a 0-dimensional subset, 
probability~1 will be assigned to properties that contain~$U$, probability~0 will be 
assigned to properties that are orthogonal to~$U$, and the remaining properties will be 
associated with nontrivial probabilities. These probabilities are uniquely determined by 
Gleason's theorem~\cite{Gleason,Pitowsky} for vector spaces with at least 3 dimensions, 
if three postulates equivalent to the following hold~\cite{Peres95}:
\bq
(A1) Attributions of the form ``System $S$ has property~$u$'' are represented by 
projectors in a complex vector space ${\cal V}(S)$.

(A2) Commuting projectors in ${\cal V}(S)$ correspond to attributions the truth values 
of which can be indicated together regardless of the truth value of each attribution.

(A3) If $P_u$ and $P_v$ are orthogonal projectors (representing attributions of $u$ and 
$v$, respectively), the probability associated with their sum $P_{uv}=P_u+P_v$, which is 
itself a projector (representing the attribution of the property spanned by $u$ and $v$), 
is the sum of the probabilities associated with $P_u$ and $P_v$.
\eq
A few remarks: (i)~The possession of the property spanned by $u$ and $v$ is 
indicated if the possession of either $u$ or $v$ is indicated, but it can also be indicated if 
neither the possession of $u$ nor the possession of $v$ is indicated. For instance, the 
electron's presence in $L\&R$ can be indicated when neither its presence in $L$ nor its 
presence in $R$ is indicated.

(ii) Suppose that the respective attributions of $u$, $v$, and $w$ are represented by 
1-dimensional projectors in a 3-dimensional vector space, that $P_w$ is orthogonal to 
both $P_u$ and $P_v$, and that $P_v$ is neither the same as nor orthogonal to 
$P_u$. Then $P_{uw}$ and $P_{vw}$ do not commute even though the corresponding 
attributions can both be true. However, if either of them is false, the other attribution 
cannot have a truth value. Hence the clause ``regardless of the truth value of either 
attribution'' in the second postulate, which is not explicit in the formulation given by 
Peres~\cite{Peres95}.

(iii) The third postulate is the classical sum rule for the probability of ``either $A$ or 
$B$,'' $p\,(A\vee B)=p\,(A)+p\,(B)$, which holds for a set of mutually exclusive events 
provided that one of them happens. Since probabilities are assigned on the proviso that 
a value be indicated, this condition is always satisfied.

A recent generalization of Gleason's theorem by P.~Busch and by Caves et 
al.~\cite{Fuchs,Busch,Cavesetal} makes the theorem applicable to 2-dimensional vector 
spaces as well. So does work by D.I. Fivel~\cite{Fivel}. The thrust of the theorem is that 
the nontrivial probabilities that are entailed by the objective indefiniteness of relative 
positions are uniquely determined by what is most inappropriately called a quantum 
``state.'' The search for a suitable probability algorithm leads to a unique 
density operator. What you seek is what you get. We sought a probability measure and 
we got a probability measure. Quantum ``states'' are probability measures.

\section{\large AGAINST ``EVOLUTION''}

While surveying several good textbooks Bell stumbled at various invocations of that 
infamous postulate according to which the quantum state of a system either 
``jumps'' from $\rho$ to $\hat{\rho}$ or ``collapses'' into an eigenstate of the observable 
that is being measured. Something is evidently wrong with this kind of narrative, but 
exactly what? For Bell the root of all evil is the special status granted to measurements 
by the authors of those books. While under ``normal'' circumstances quantum states 
evolve deterministically, measurements are said to cause them to change 
discontinuously. Yet the elimination of all references to measurements qua ``piddling 
laboratory operations'' does not solve the problem posed by the existence of two kinds of 
evolution. Measurement outcomes qua property-indicating facts are an integral part of 
QM. The properties of the quantum world supervene on the facts; they need to be 
indicated in order to exist. The root of the problem is not the extrinsic nature of the 
values of quantum-mechanical observables but the notion that a quantum state is 
something that can evolve.

Fuchs and Peres stress that
\bq
the time dependence of the wavefunction does not 
represent the evolution of a physical system. It only gives the evolution of our 
probabilities for the outcomes of potential experiments on that system.~\cite{FuPer}
\eq
The wave function being solely a ``compendium of probabilities,'' what holds for the 
time dependence of probabilities must also hold for the time dependence of the 
quantum states that define them. If probabilities evolve, so do quantum states. On this 
account, quantum states are things that exist and change in time, for only such things 
can be said to evolve.

By denying that the evolution of a quantum state represents the evolution of a physical 
system, Fuchs and Peres avoid the various ``quantum paradoxes'' entailed by realistic 
construals of quantum states. However, the notion that probabilities (and hence 
quantum states) nevertheless evolve entails inconsistencies of its own. For instance, 
the authors state that ``no wavefunction exists either before or after we conduct an 
experiment.'' This is correct. The time on which a wave function depends is 
the time at which an observable has an indicated value, either in the actual world or 
counterfactually, on the false assumption (but nevertheless on the assumption) that a 
value is indicated. Hence no wave function ``exists'' at a time that is not the time of an 
actual or counterfactual measurement. But this is clearly inconsistent with the notion 
that quantum states evolve.

Again, Fuchs and Peres stress that ``collapse is something that happens in our 
description of the system, not to the system itself.'' But if collapse does not happen to the 
system, a ``description'' of the system in which it happens cannot be a (correct, valid, 
acceptable) description! This distinction between knowledge or a description on the one 
hand and the thing known or described on the other has always baffled me. How can 
there be ``an abstract quantum physical description'' if ``[t]here is no quantum 
world''~\cite{Petersen} answering that description? What does such a description 
describe? Knowledge? If so, knowledge about what? I fear that the commendable 
objective of these authors---to deflate realistic construals of the quantum state---will not 
be achieved by the self-contradictory and philosophically defunct stratagem of 
driving a wedge between knowledge and the known. What we need is a correct 
description, and once we have it, we can stop talking about our description and instead 
talk about the thing described.

If collapse does not happen to the system, it cannot happen in a valid description of the 
system. So what makes it seem to happen? Nothing but the erroneous notion that 
quantum states (or the probabilities they assign) evolve. What evolves, evolves in 
relation to a detached, intrinsically differentiated temporal background. An evolving 
quantum state, if it existed, would be a function of a succession of self-existing instants. 
It would exist at every ``moment of time.'' Yet, so Fuchs and Peres assure us, no wave 
function exists at a time that is not the time of an actual (or counterfactual) 
measurement. Nor, as a matter of fact, does a wave function exist at the time of a 
measurement, for the probability for something to happen at a certain time is not 
something that exists at that time, any more than the probability for something to be 
found in a region $V$ is something that exists in~$V$. Probabilities (and hence quantum 
states) do not exist at {\it any\/} time. {\it A fortiori\/} they are not things that evolve, 
whether deterministically or discontinuously.

The question, therefore, is not whether quantum states are states of Nature or states of 
knowledge. This question only arises if quantum ``states'' are conceived as 
{\it states\/}---something that exists in time and evolves. If quantum ``states'' were 
states, Fuchs and Peres would be right: They could only be states of knowledge. The idea 
that quantum states are states of knowledge is therefore a direct consequence of the 
notion that quantum states evolve. Quantum realists and the proponents of epistemic 
interpretations are equally wrong, inasmuch as they both take for granted that the 
temporal referent of a quantum state is a detached, intrinsically differentiated temporal 
background. The question is not whether $\rho(x,t)$ represents a state of this or that 
kind---ontological or epistemic. The question is what the true referents of $t$ and $x$ 
are if they do not refer to a detached, intrinsically differentiated spatiotemporal 
background. The real issue is not the kind of reality that {\it quantum states\/} possess 
but the kind of reality that {\it the spatial and temporal referents of quantum states\/} 
possess. Where space is concerned we already know the answer: a relative and 
contingent kind. It will come as no surprise that the same holds true for time.

A quantum state is a generic probability measure. It depends on a set of indicated 
properties, which constitute the factual basis on which probabilities are assigned. The 
generic measure defines specific measures, which depend on specific observables and on 
the time $t$ of measurement. In the case of a position measurement on~$O$, the specific 
measure $p\,(V_i,t)$ further depends on a partition of space into regions~$V_i$. Such a 
region exists for $O$ if and only if the proposition ``$O$~is in~$V_i$'' has a truth value. 
This is the case if and only if a truth value is indicated, and this is possible only if the 
region is realized (made real) by being an intrinsic property of a macroscopic detector.

The regions $V_i$ thus are not self-existent and intrinsically distinct ``parts of space.'' 
They are realized by macroscopic detectors. There is no such thing as a detached, 
intrinsically differentiated spatial background. Space consists of the indicated spatial 
properties of material objects, and these supervene on the self-indicating spatial 
properties of macroscopic objects. By the same token, the time on which $p\,(V_i,t)$ 
depends is not a self-existent ``moment of time.'' There is no such thing as a detached, 
intrinsically differentiated temporal background. Times, too, are extrinsic. Time consists 
of the times that are indicated by macroscopic clocks (that is, by the positions of 
macroscopic objects that are suitable for indicating time). Times exist to the extent that 
they are indicated, and a time~$t$ exists for a system~$S$ if and only if it is the time of 
a ``measurement'' (that is, the indicated time at which an observable on~$S$ possesses an 
indicated value).

Thus when Fuchs and Peres say that ``no wavefunction exists either before or after we 
conduct an experiment,'' which suggests that a wave function exists at the time of 
measurement, what they really mean (or ought to mean) is that the only {\it 
times\/} that exist for a quantum system are the indicated times at which the system has 
indicated properties. This is why it makes no sense to speak of an ``evolution.'' The word 
``evolution'' is appropriate neither for the time dependence of a generic probability 
measure that is based on a ``preparation'' (that is, on a given set of property-indicating 
facts), nor for the replacement of a generic probability measure based on one set of facts 
by a generic probability measure based on a different set of facts.

\section{\large QUANTUM MECHANICS IS COMPLETE}

QM is about probabilities. Until 1926, when this was first understood by Max Born, all 
probabilities known in physics were concessions to our {\it ignorance\/} of some of the 
relevant data, and therefore essentially {\it subjective\/}. This is no longer the case. Born 
probabilities are assigned by fundamental physical laws, on the basis of particular sets of 
facts, in a way that does not involve anyone's knowledge or ignorance of the relevant 
facts. A clear distinction has to be maintained between physical laws and the uses we 
make of them. We may use them to make statistical predictions on the basis of whatever 
facts we were able to gather. In doing so we single out a particular set of facts. We can 
make use only of known facts. But the laws of QM do not single out any particular set of 
facts. They assign nontrivial probabilities to the possible values of {\it any\/} observable 
and on the basis of {\it any\/} set of relevant facts. These probability assignments are 
fundamental physical laws. They are as objective as the indefiniteness of which they are 
the formal expression, and they are objective in the same sense in which the laws of 
classical physical would be objective if they were true. QM provides as much a model of 
a free-standing world as classical physics did. The only difference is that the classical 
model is deterministic while the quantal model incorporates an objective 
indefiniteness in the form of an objective probability algorithm.

Why are we not satisfied with this? Probably because we harbor the illusion that a 
deterministic description provides a complete explanation. That this is an illusion was 
pointed out two and a half centuries ago by Immanuel Kant. Newton knew that he had 
{\it described\/} gravity but was unable to {\it explain\/} it. Kant made it clear why:
\bq
That the possibility of fundamental forces should be made conceivable is a completely 
impossible demand; for they are called fundamental forces precisely because they cannot 
be derived from any other force, i.e. they cannot be conceived.~\cite{Kant}
\eq
This is as true today as it was in Kant's time. A fundamental physical theory is, by 
virtue of being fundamental, an inexplicable {\it description\/} of correlations among the 
goings-on in the world, regardless of whether these are instantaneous or 
retarded, local or nonlocal, deterministic or probabilistic. The question, therefore, is not 
whether QM affords a complete explanation but only whether it affords a complete 
description. And that it does. A description that includes everything that is indicated is 
complete. If anything is incomplete it is the physical world itself, but this is incomplete 
only in relation to an imaginary spatiotemporal background that is more differentiated 
than the physical world.

The question of why the world answers the description that it does is beyond the reach 
of physics. So is the question of why it does answer a description that owes so 
much to logic, probability theory, and pure mathematics, and comparatively little to 
empirical input. ``The most incomprehensible thing about the world is that it is 
comprehensible,'' Einstein remarked~\cite{Schilpp1949}. Other physicists have voiced 
similar astonishment. To countless philosophers since Plato the comprehensibility of the 
world has suggested an affinity of our minds with the Power or Principle responsible for 
the existence of the physical world. I won't contradict anyone who feels the necessity of 
conceiving of such a Power or Principle. But I would insist that this necessity did not 
arise with the discovery of QM; it arose when philosophers first began to think about 
the reality of concepts~\cite{vW}.

\section{\large CONCLUSION}

In the opening paragraph of a recent review article by F. Lalo\"e~\cite{Laloe} two 
interpretational issues are mentioned by way of example. One of them is the question, 
does $\ket\psi$ ``describe the physical reality itself, or only some partial knowledge that 
we might have of this reality?'' Further on much the same question is identified as one of 
the key issues:
\bq
To what extent should we consider that the wave function describes a physical system 
itself (realistic interpretation), or rather that it contains only the information that we 
may have on it (positivistic interpretation)\dots?
\eq
This false disjunction leaves no room for a successful resolution of the interpretational 
issues raised by QM. It has its roots in an illegitimate projection of an imaginary 
spatiotemporal background into the physical world. The wave function itself neither 
describes a physical system nor does it contain {\it only\/} the information that {\it we\/} 
may have. An objective probability measure, it is the proper expression of an objective 
indefiniteness. Its failure to describe the state of the system is not a shortcoming but a 
consequence of the nonexistence of a detached temporal background that would allow us 
to assign a state to the system at every ``moment of time.'' The temporal referent of a 
quantum state (in the Schr\"odinger picture) is not a self-existent instant but the 
indicated time of possession of an indicated property. The spatiotemporal properties of 
the world supervene on the domain of facts, which is constituted by the positions of 
macroscopic objects.

The second part of Lalo\"e's article, entitled ``Difficulties, Paradoxes,'' begins as follows:
\bq
We have seen that, in most cases, the wave function evolves gently, in a perfectly 
predictable and continuous way, according to the Schr\"odinger equation; in some cases 
only (as soon as a measurement is performed), unpredictable changes take place, 
according to the postulate of wave packet reduction.
\eq
Once quantum state evolution is taken for granted one is confronted with two distinct 
modes of evolution. This has not only led to a host of inane paradoxes but also elicited 
such inappropriate responses as the claim that ``state reductions\dots are nothing but 
mental processes''~\cite{ESW}. If quantum states did evolve, if there were such things as 
state reductions, or if state reductions were processes, then I suppose they would have to 
be regarded as mental processes. But none of the above antecedents is true. Although the 
literature is infested with statements suggesting the contrary (for instance, ``coherent 
superpositions tend to constantly propagate toward the environment''~\cite{Laloe}), the 
time dependence of the density matrix has nothing to do with any process, whether 
physical or mental. ``Logically it is\dots clear that this problem will never be solved by 
invoking any process that is entirely contained in the linear Schr\"odinger equation,'' 
Lalo\"e writes. It would be more appropriate to say that this problem will never be 
solved by {\it any\/} process. Nor is any process contained in the Schr\"odinger equation. 
Unless we learn to talk about probability measures in a language that is suitable for 
probability measures, I fear that making sense of QM will remain a distant goal.

But perhaps the situation is not all that bleak. The first interpretation discussed by 
Lalo\"e, a sort of common ground he calls the ``correlation interpretation'' (CORI), 
emphasizes the correlations between ``successive results of experiments.'' Significantly, 
in this interpretation ``no conflict of postulates takes place\dots\ no paradox can be 
expressed in terms of correlations.'' The trouble starts when the CORI is conjoined with 
a detached, intrinsically differentiated spatiotemporal background. This is what has 
engendered all those interpretations whose
\bq
general purpose always remains the same: to solve the problems and questions that are 
associated with the coexistence of two postulates for the evolution of the state 
vector.~\cite{Laloe}
\eq
Objections to the CORI are of two kinds. Those of the first kind are either misconceived 
or readily parried. The others can be defused by going beyond the CORI without 
invoking a detached spatiotemporal background. One objection, directed against the 
perceived undue emphasis on experiments, is easily rebutted by drawing the appropriate 
distinction between experiments and property-indicating facts. The emphasis on the 
latter---the supervenience of properties on facts---is a direct consequence of an objective 
indefiniteness in the physical world. Another perceived shortcoming---that the CORI 
does not tell us ``which sequence [of possible measurement results] is realized in a 
particular experiment''---is not a shortcoming at all but a consequence of this objective 
indefiniteness.

The objection that the CORI ``shows no interest whatsoever in questions related to 
physical reality as something `in itself'\,'' is of the second kind. (So is the objection that 
``the boundary between the measured system and the environment of the measuring 
devices is flexible,'' which has been dealt with in Sec.~3.) Nobody can be forced to show 
interest in such questions. Unfortunately some proponents of the CORI go further by 
declaring that QM is inconsistent with a free-standing physical reality. What QM is 
actually inconsistent with is the notion of a detached, intrinsically differentiated 
spatiotemporal background. Once this is rejected, a consistent conception of a 
free-standing quantum world is perfectly possible, as this article attempted to 
demonstrate. The quantum world consists of the positions of macroscopic objects and of 
whatever is indicated by them. The 
dependence that has proved such a stumbling block to understanding QM is not a 
dependence on anything external to the quantum world but (i)~an internal, mutual 
dependence of the positions of macroscopic objects and (ii)~the supervenience on these 
positions of the remaining properties of the quantum world, which is a direct 
consequence of the indefiniteness of the quantum world.

A possible objection to the emphasis on {\it positions\/} has been met by Lalo\"e himself 
(if one reads ``property-indicating facts'' for ``experiments'' and ``indicated'' for 
``measured''):
\bq
One can easily convince oneself that\dots what is measured in all experiments is basically 
the positions of particles or objects (pointers, etc.), while momenta are only indirectly 
measured.
\eq
One final question: Why does the present article, including its title, single out epistemic 
interpretations, considering that realistic interpretations of quantum {\it states\/} are 
even farther off the mark? The answer is that the absurd consequences of the latter 
interpretations have 
been adequately highlighted by the proponents of the former, while the proponents of 
the latter are unable to adequately highlight the errors of the former inasmuch as they 
share them.


\begin{thebibliography}{99}
\bibitem{Relationalism}
The view that no position exists unless it is possessed accords with relationalism, 
the doctrine that space (and time) must be thought of as a family of spatial (and 
temporal) relations holding among the material constituents of the universe. The 
view that no position is possessed unless its possession is indicated goes beyond 
previous versions of relationalism (such as that advocated by Leibniz), which 
treat spatial relations as intrinsic.
\end{thebibliography}
\end{document}